\documentclass[conference]{acmsiggraph}

\TOGonlineid{0032}
\TOGvolume{0}
\TOGnumber{0}
\TOGarticleDOI{1111111.2222222}
\TOGprojectURL{}
\TOGvideoURL{}
\TOGdataURL{}
\TOGcodeURL{}

\usepackage{subfig}    
\usepackage{subfloat}  
\usepackage{amsfonts}  

\title{History-free Collision Response for Deformable Surfaces}

\author{Juntao Ye\thanks{e-mail:juntao.ye@ia.ac.cn}\\Institute of Automation, Chinese Academy of Sciences}
\pdfauthor{J. Ye}


\begin{document}

\teaser{
  \subfloat[]{
  \includegraphics[width=.40\linewidth]{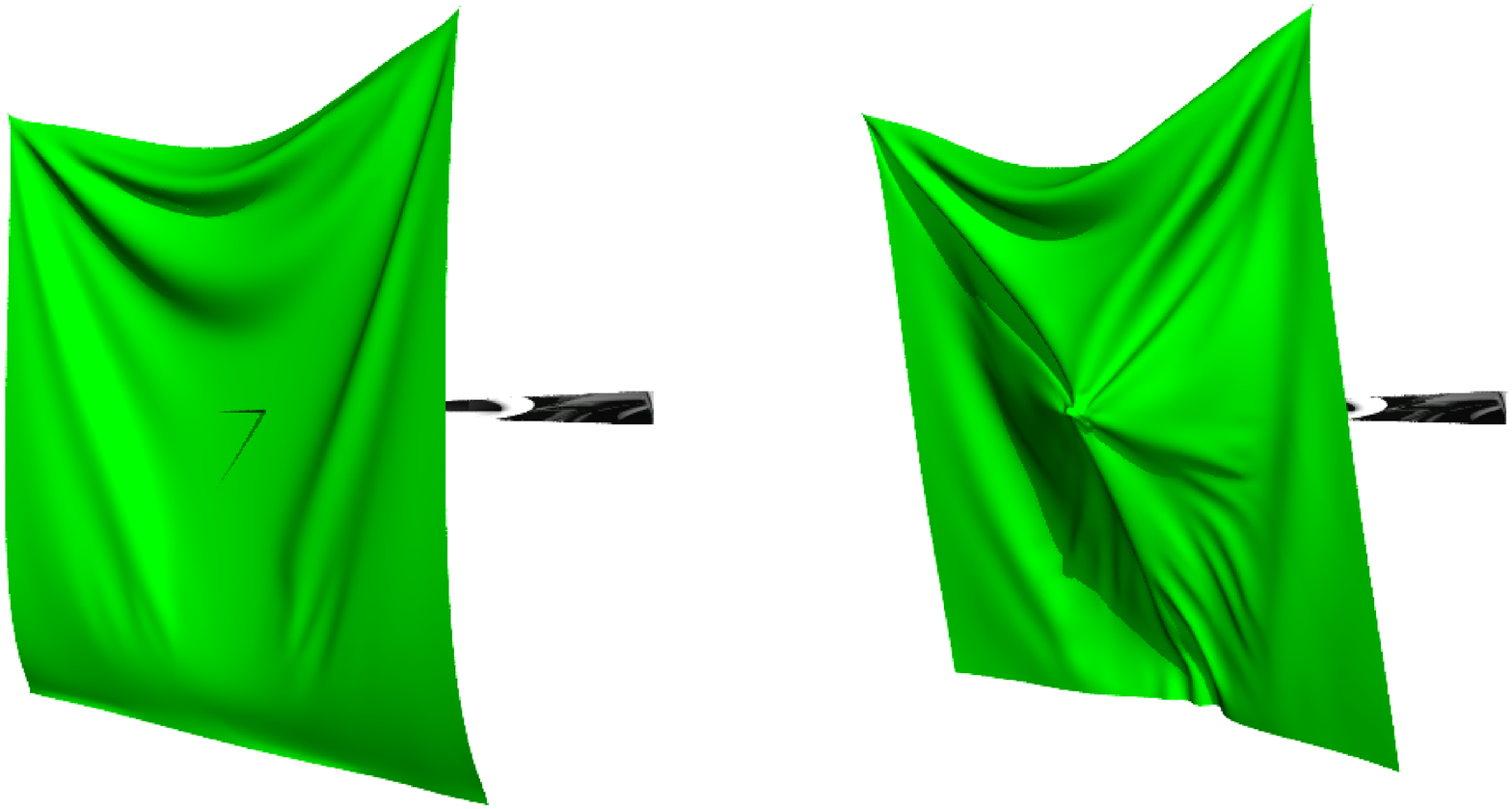}}
  \subfloat[]{
  \includegraphics[width=.60\linewidth]{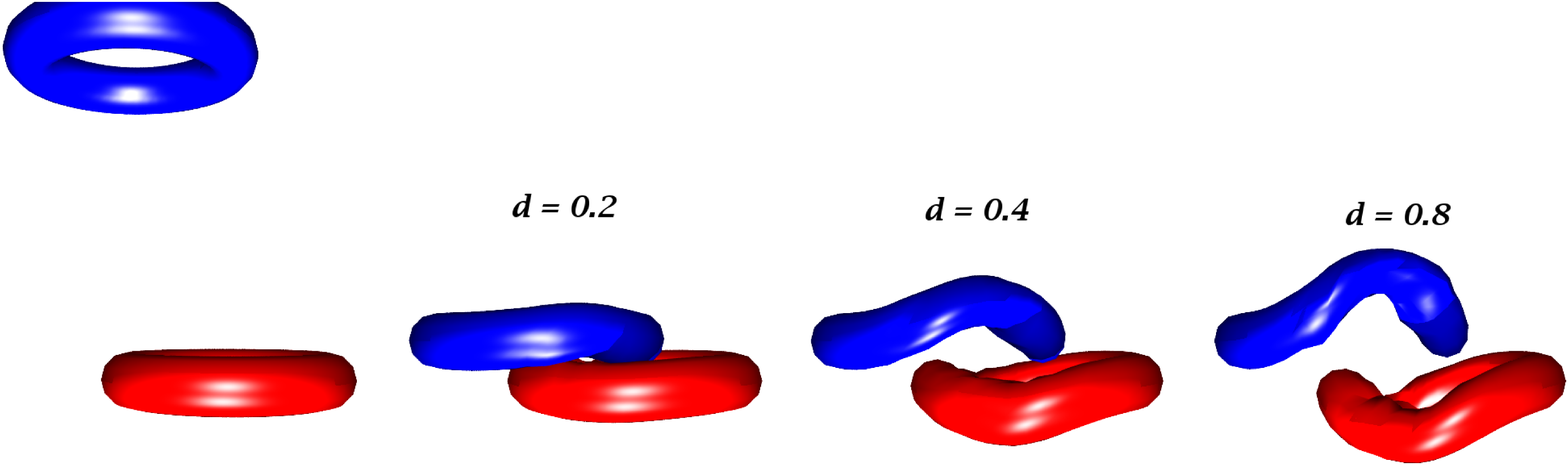}}
  \caption{(a) An existing penetration configuration is recovered with our DCD-based response method.
           (b) The blue torus is falling on to the red one. By setting different post-response distances,
               a range of behaviors from inelastic to elastic collision can be simulated.}
  \label{fig_teaser}
}

\maketitle

\begin{abstract}
Continuous collision detection (CCD) and response methods are widely adopted in dynamics simulation of deformable models.
They are history-based, as their success is strictly based on an assumption of a collision-free state at the start of each time interval.
On the other hand, in many applications surfaces have normals defined to designate their orientation (i.e. front- and back-face),
yet CCD methods are totally blind to such orientation identification (thus are orientation-free).
We notice that if such information is utilized, many penetrations can be untangled.
In this paper we present a history-free method for separation of two penetrating meshes,
where at least one of them has clarified surface orientation.
This method first computes all edge-face (E-F) intersections with discrete collision detection (DCD),
and then builds a number of penetration stencils.
On response, the stencil vertices are relocated into a penetration-free state, via a global displacement minimizer.
Our method is very effective for handling penetration between two meshes,
being it an initial configuration or in the middle of physics simulation.
The major limitation is that it is not applicable to self-collision within one mesh at the time being.
\end{abstract}

\begin{CRcatlist}
  \CRcat{I.3.3}{Computer Graphics}{Three-Dimensional Graphics and Realism}{Display Algorithms}
  \CRcat{I.3.7}{Computer Graphics}{Three-Dimensional Graphics and Realism}{Radiosity};
\end{CRcatlist}

\keywordlist


\TOGlinkslist


\copyrightspace

\section{Introduction}
\label{sec:introduction}

Collisions are often unavoidable in modeling deformable surfaces such as cloth.
The majority simulation systems adopt CCD to predict impending collisions,
then attempt to prevent them from happening by altering particles' velocities.
The success of CCD-based response method relies on a hard constraint: a collision-free state for
not only the initial configuration but also the start of every time interval in the simulation.
This requirement makes CCD methods to be called history-based.
A major advantage of CCD methods is that they do not rely on surface orientation information
to decide whether a geometric primitive is located on the incorrect side of a surface or not.
This is very important for surfaces (e.g. a sheet model) which typically do not present orientation information.
However, there also exists surfaces which do exhibits orientation information, either explicitly or implicitly.
For example, in a polygonal sphere model, the polygon normals usually point to the outside of the sphere;
another example is the simulation of a dressed avatar, in which the normals of the body surface also points to the outside
so that no cloth vertices are allowed to go inside the body.
We call them {\it oriented surfaces}.
CCD methods, designed to handle generic collision context which could involve either oriented or un-oriented surfaces,
are totally blind to such useful orientation information.
Nevertheless, normals are indispensable ingredients in CCD and their directions for colliding polygons
are so picked as to be consistent with the negative approaching vertex velocities.
In this sense orientations of those polygons are derived dynamically from the history.
For oriented surfaces, this derivation seems redundant.
An interesting question is, if orientation is available, will it be more helpful
than history information?

Our goal is to put forward a new method that takes advantage of
surface orientation if it is available, and no longer relies on the history information.
Different from CCD as prevention-based methods, our method embraces the {\it repair} strategy --
penetrations are allowed to occur but will be detected and penetrating objects will be separated.
It employs {\itshape discrete collision detection} (DCD) so it is history-free.
DCD only finds out edge-face (E-F) intersections for two given meshes, thus is faster than CCD for two points:
bounding volumes are more tight-fitting and no cubic solver is needed.
Please note that at the time being our method is capable of handling collisions
between two different meshes, at least one of them being oriented.
How to extend our method to handle self-collision is still under investigation.

\section{Related work}
\label{sec:related}

A number of research groups have worked on the CCD method.
\cite{provot97collision} proposed a cubic solver for detecting penetration of two elastic moving triangles.
\cite{bridson02robust} presents a comprehensive way of preventing intersections to occur.
CCD once suffered a serious problem that the cubic polynomial solver is vulnerable to round-off error.
This problem was reported to be solved not long ago \cite{brochu12exact}.
CCD helped the asynchronous contact model \cite{Harmon09asynchronous,Ainsley12specu} to produce very realistic effects.
In CCD, every detected collision event carries a time tag, but they are often ignored,
as resolving all the events strictly according to their time order can be too slow to halt the simulation.
Thus all events are often assumed to occur at exactly the same time instant and
simultaneous response is adopted \cite{Harmon08robust}.
This demands the time interval for CCD to be of moderate size.

CCD will suffer even serious problem in applications in which an intersection-free initial state is impossible,
or the simulation context is subject to external constraints
forcing the deformable surface into illegal states for a number of consecutive steps.
In these situations if some vertices end up on the back-face side of another surface,
they will be ``remembered" as not, and the simulator will work
hard to keep them on the wrong side forever.
Therefore, a ``repair" mechanism is needed for robust simulation.
Baraff {\itshape et. al} \cite{baraff03untangling} were probably the first to address the necessity of
collision correction in complex simulation environment, and put forward a partial solution.
Their method is limited to closed regions and cannot handle intersection involving mesh boundary.
Wicke {\itshape et. al} \cite{wicke06untangling} further investigated this problem by taking consideration of boundary collisions.
They analyzed all possible collision configurations and presented a classification criterion for intersection contours.
An intersection contour minimization (ICM) method \cite{volino06contour} was proposed.
While their algorithm are capable of handling certain very complex situations,
it also fails for some other situations.
The failure or slow convergence of ICM is due to ambiguous normal direction used according to \cite{ye12contour},
therefore it is only capable of handling boundary-involved cases
but is not good at closed contours.
Narain {\itshape et. al} \cite{Narain2012AAR} encountered the interpenetration problem while re-meshing cloth
to enrich details, and the penetrating vertices are moved back in a manner quite similar to our method in this paper.

\begin{figure}[!htbp]
  \centering
  \subfloat[]{
  \includegraphics[width=.24\linewidth]{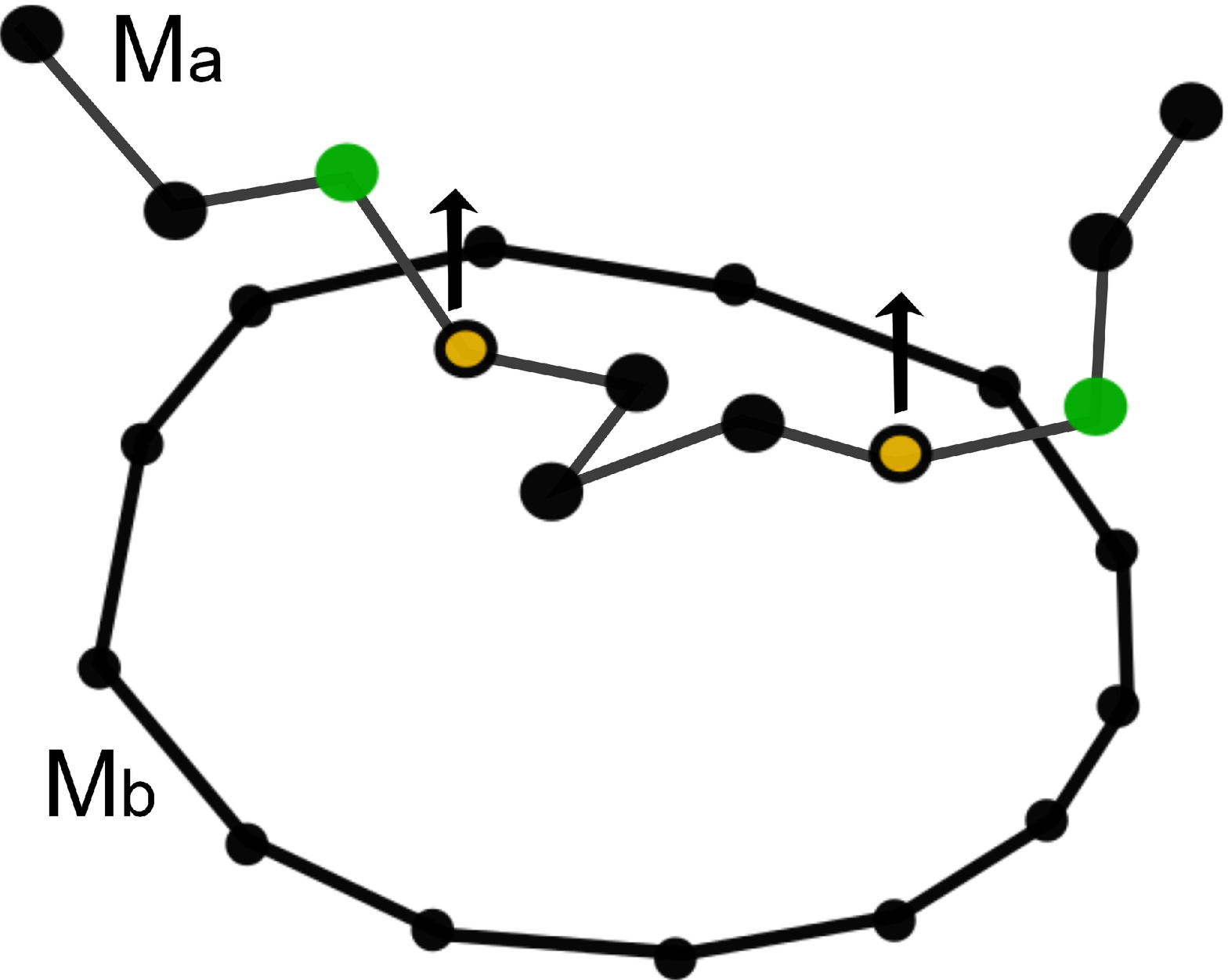}}
  \subfloat[]{
  \includegraphics[width=.24\linewidth]{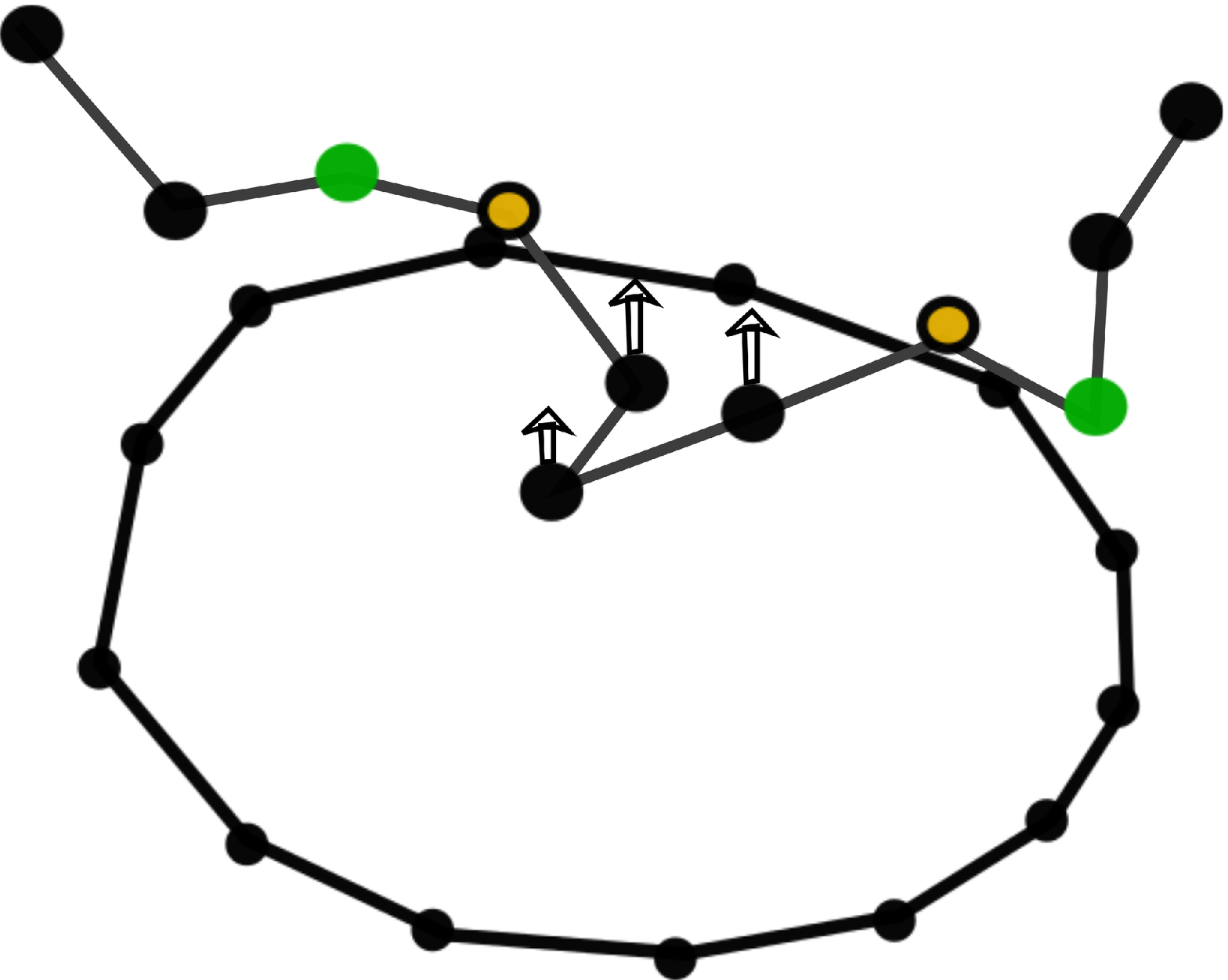}}
  \subfloat[]{
  \includegraphics[width=.24\linewidth]{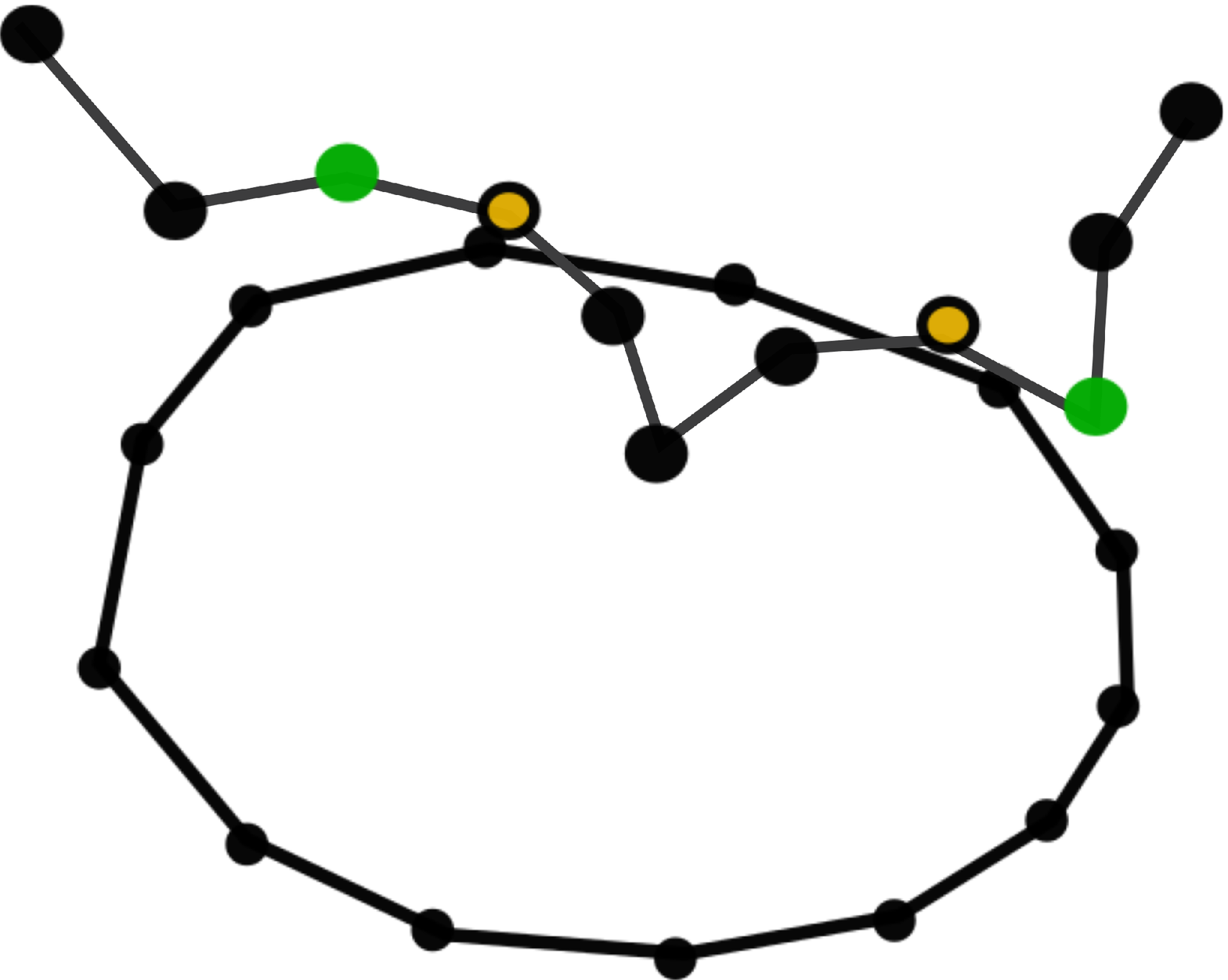}}
  \subfloat[]{
  \includegraphics[width=.24\linewidth]{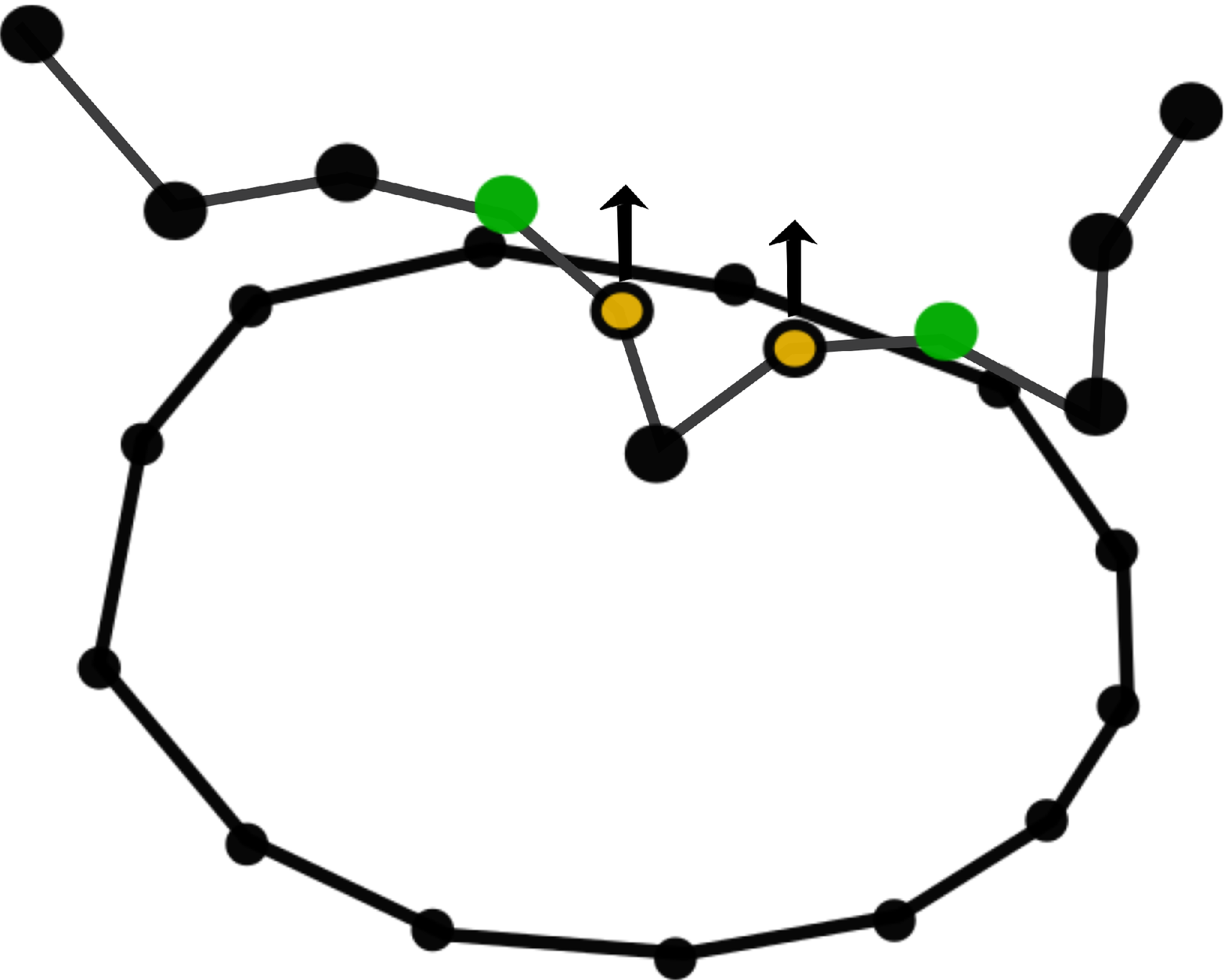}}\\
  \subfloat[]{
  \includegraphics[width=.24\linewidth]{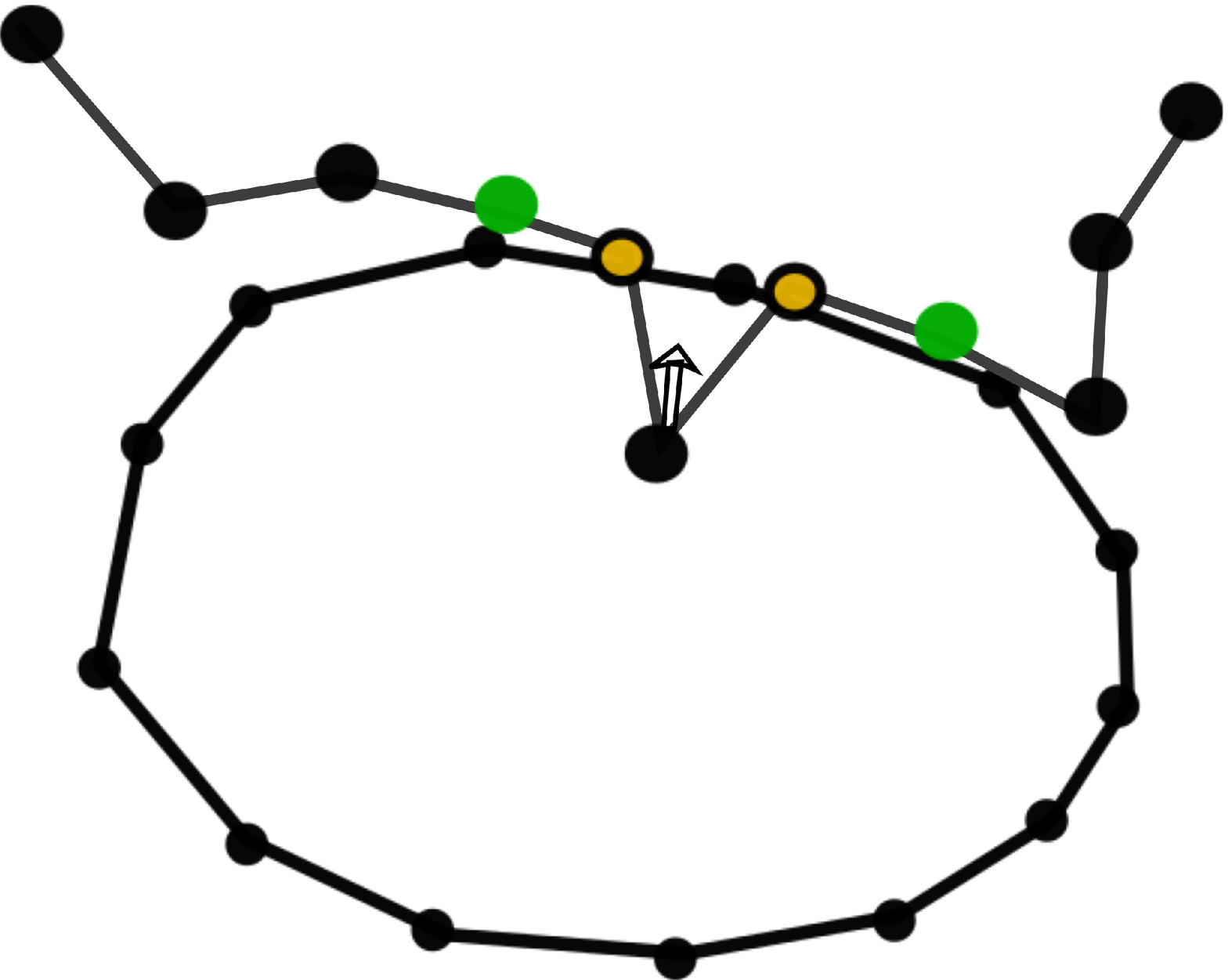}}
  \subfloat[]{
  \includegraphics[width=.24\linewidth]{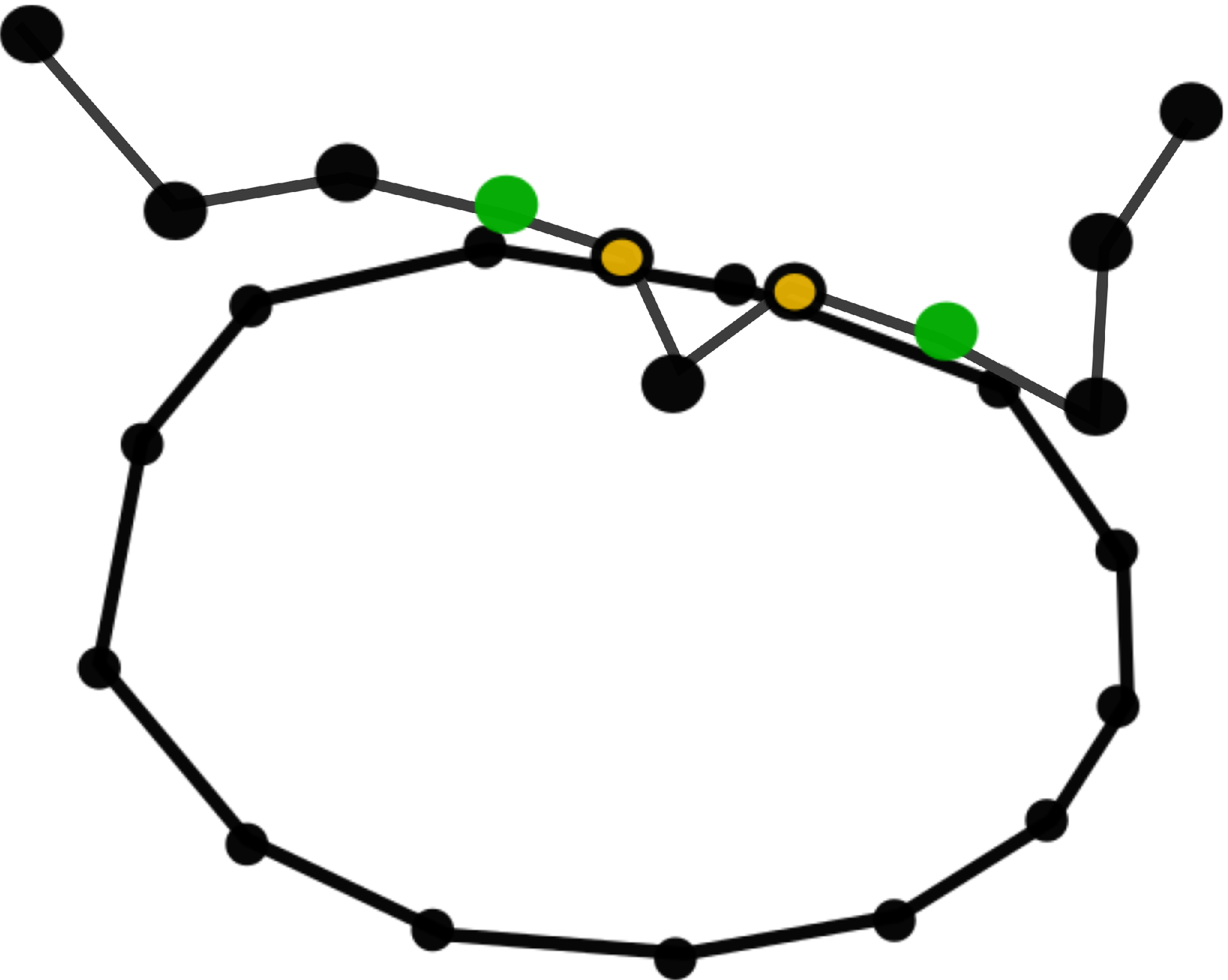}}
  \subfloat[]{
  \includegraphics[width=.24\linewidth]{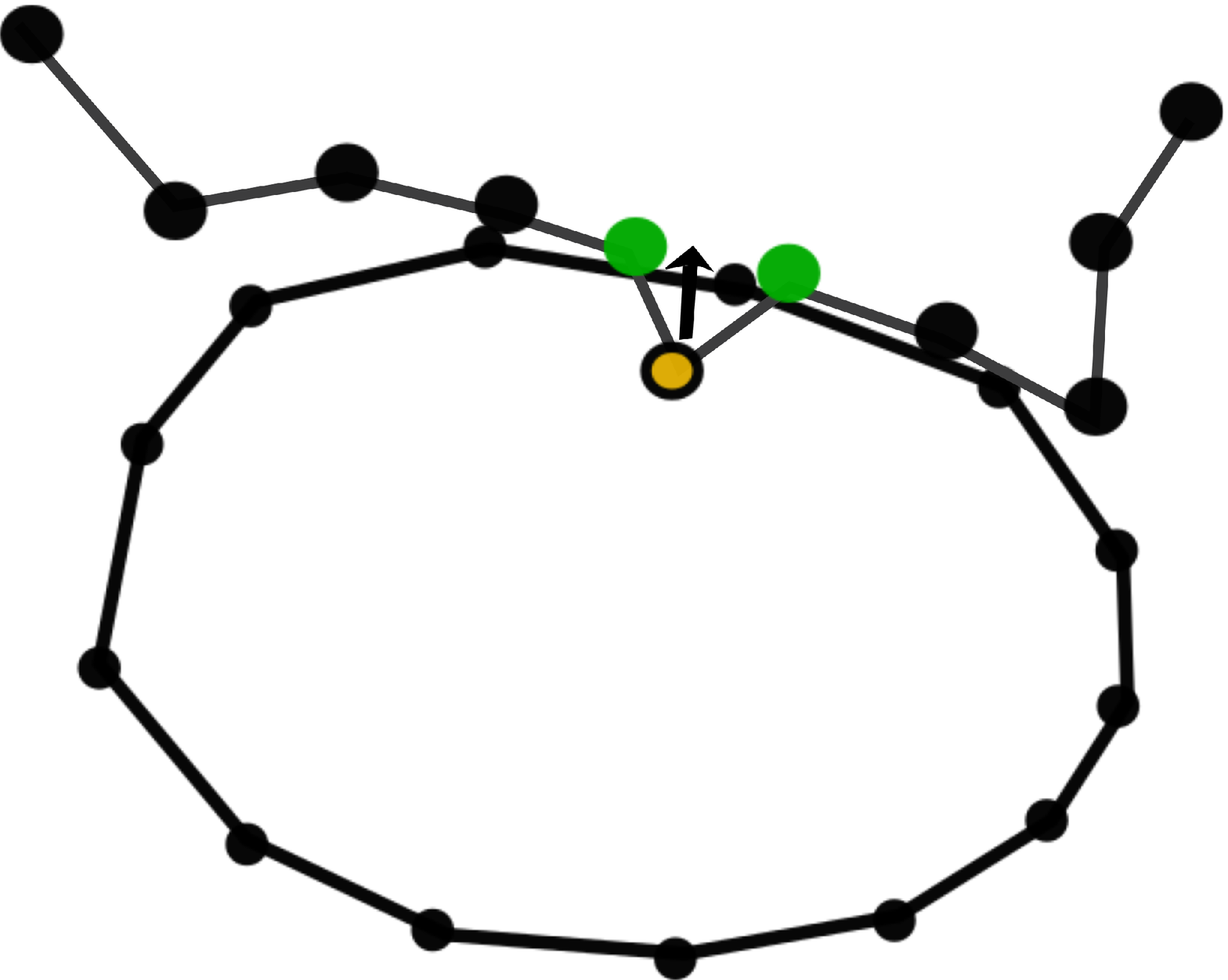}}
  \subfloat[]{
  \includegraphics[width=.24\linewidth]{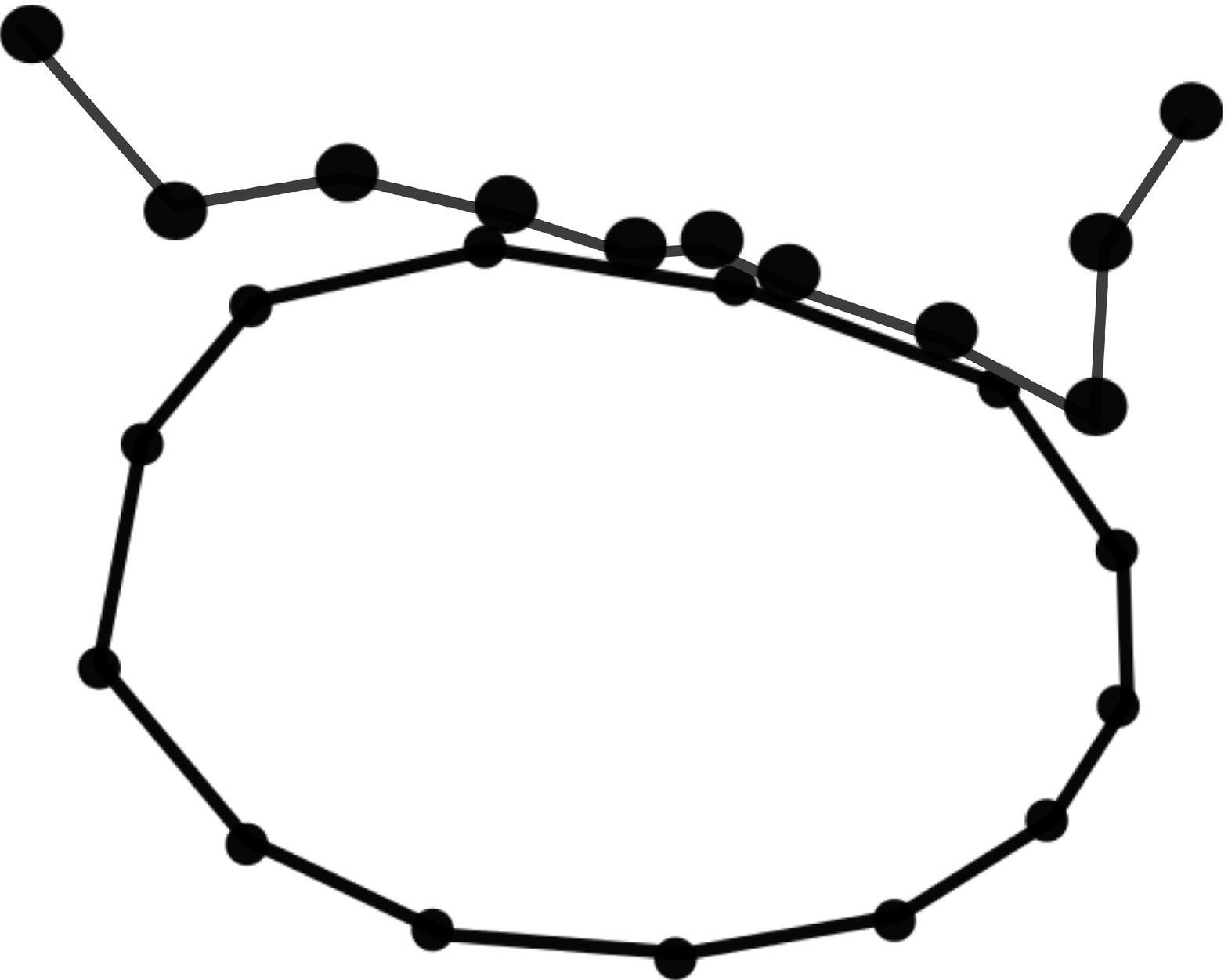}}
  \caption{\small The 2D illustration of how our method works.
           Mesh $M_b$ is oriented surface of a solid object, and $M_a$ is un-oriented deformable surface.
           In the penetration configuration:
           vertex of an intersecting edge is marked {\it green} if it is in legal state,
           or marked {\it yellow} if it is in illegal state;
           a vertex is marked {\it black} if it is not on any intersecting edge.
           }
  \label{fig:algoverview}
\end{figure}

\section{Algorithm overview}

We show how our method works by a 2D illustration of a penetration configuration between one
oriented surface $M_b$ and one un-oriented surface $M_a$ in Fig.~\ref{fig:algoverview}.
The illegally located vertices are iteratively detected and then relocated.
First the DCD algorithm finds all E-F intersections,
in which the face is from the oriented face and the edge is from the un-oriented one.
Then the legality of each end-vertex of these edges is then determined according to the front-face identification:
legal vertices are marked green and illegal vertices are yellow.
For those edges that do not intersect any faces, the legality of their end-vertices can not be determined and
these vertices are marked black (Fig.~\ref{fig:algoverview}a).
The illegal (yellow) vertices are to be relocated to legal positions via our {\it minimum positional displacement}
algorithm (Fig.~\ref{fig:algoverview}b), and also the displacement vector will be diffused to the vicinity (Fig.~\ref{fig:algoverview}c).
These yellow vertices become green in the next round of DCD (Fig.~\ref{fig:algoverview}d).
Repeat this process until there is no E-F intersections (Fig.~\ref{fig:algoverview}e -\ref{fig:algoverview}h).
Our algorithm is so designed that the yellow vertices are moved in parallel, along the shortest paths,
to reach penetration-free state.

\section{Minimum positional displacement}
\label{sec:Model}

In this paper, the surfaces are described as polygonal (particularly triangular) meshes.
The entire mesh, at any time instant, can be viewed as a point in a high-dimensional space.
For an {\it n}-vertex mesh, we denote this point as
${\bf x}(t) = [{\bf x}_0(t), {\bf x}_1(t), \cdots, {\bf x}_{n-1}(t)] \in \mathbb{R}^{3n}$.
At any time instant, surface intersections consist of a number of face-face (F-F) intersections.

\begin{figure}[!htbp]
  \centering
  \subfloat[]{
  \includegraphics[width=0.25\linewidth,height=.24\linewidth]{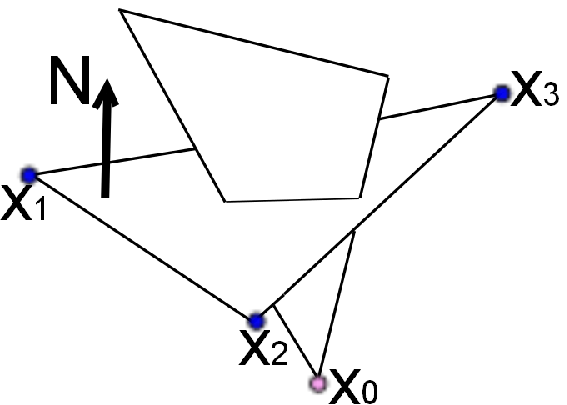}}
  \subfloat[]{
  \includegraphics[width=0.25\linewidth,height=.24\linewidth]{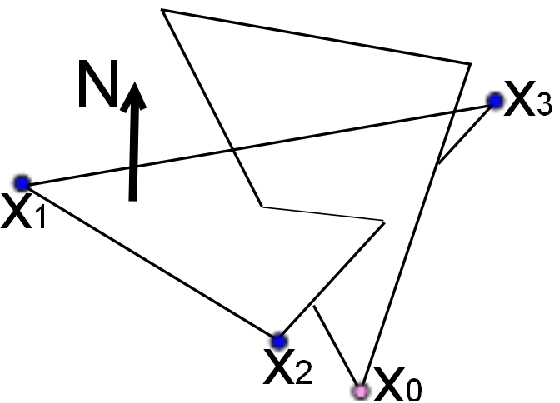}}
  \subfloat[]{
  \includegraphics[width=0.25\linewidth,height=.24\linewidth]{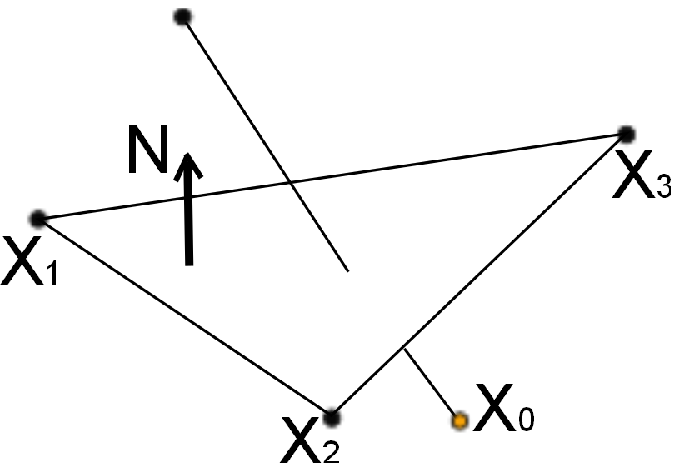}}
  \subfloat[]{
  \includegraphics[width=0.25\linewidth,height=.24\linewidth]{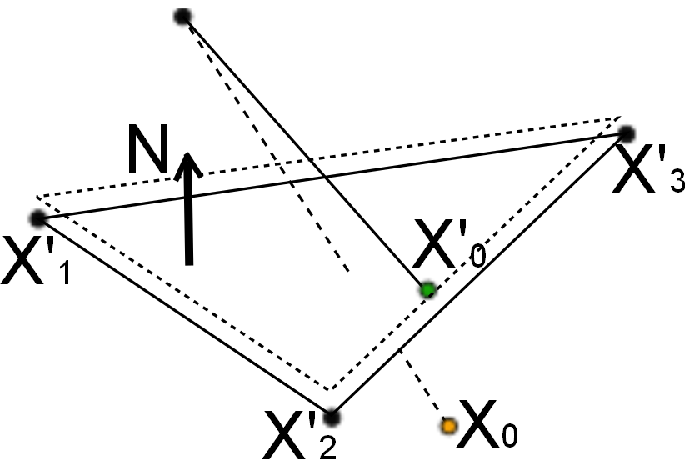}}
  \caption{(a)(b) are two types of face-face intersection;
           (c) four vertices \{${\bf x}_0$, ${\bf x}_1$, ${\bf x}_2$, ${\bf x}_3$\} form a V-F penetration stencil, and in
           (d) penetration is resolved by relocating four vertices so that ${\bf x}'_0$ is above the triangle plane.}
  \label{fig:penetrationStencil}
\end{figure}

Fig.~\ref{fig:penetrationStencil}(a)(b) shows two cases of F-F penetration,
in which one face has a normal designating its front-face orientation and the other is un-oriented.
In both cases, vertex ${\bf x}_0$ is located illegally on the back-face side of
${\bf x}_1 {\bf x}_2 {\bf x}_3$ (Fig~\ref{fig:penetrationStencil}(c)).
If we could relocate these vertices so that ${\bf x}_0$ goes to the other side of the face (Fig~\ref{fig:penetrationStencil}(d)),
this penetration will be resolved.
As we can see an F-F penetration can always be decomposed into two E-F penetrations,
the E-F penetration then becomes the elementary penetration to be resolved.
We borrow a terminology from \cite{Harmon08robust} and call the quadruple \{${\bf x}_0$, ${\bf x}_1$, ${\bf x}_2$, ${\bf x}_3$\}
a {\it penetration stencil}.
A stencil has three vertices coming from an oriented face, and one vertex from an edge, be it from an oriented surface or not.
The F-F penetration in Fig~\ref{fig:penetrationStencil}a has two E-F penetrations, leading to two stencils (yet they are identical).
The two E-F penetrations in Fig~\ref{fig:penetrationStencil}b result in only one stencil,
as one face has no orientation.
Please note if both faces are orientated we interchangeably regarded one as oriented and other as un-oriented in the iterations.

{\bf Penetration Constraints.}
We view each E-F penetration as the violation of a scalar-valued constraint function, $D({\bf x})$,
with $D({\bf x}) < 0$ whenever ${\bf x}$ is in a penetration configuration.
For V-F penetration the constraint function can be considered as
a {\it signed distance} from vertex ${\bf x}_0$ to face ${\bf x}_1 {\bf x}_2 {\bf x}_3$:
\begin{equation}
D({\bf x}) = {\bf n} \cdot [{\bf x}_0 - (\alpha_1 {\bf x}_1 + \alpha_2 {\bf x}_2 + \alpha_3 {\bf x}_3)]
\label{eqn_post01}
\end{equation}
where ${\bf n}$ is the triangle normal, and $\alpha_1, \alpha_2, \alpha_3$ are the barycentric coordinates
of any point on the plane spanned by the triangle.
As will be clarified later on, we choose $\alpha_{1,2,3} = 1/3$, anticipating that the three vertices are
undergoing the same displacement vector during the relocation.

While each constraint depends on only four vertices,
it can be differentiated with respect to ${\bf x}$ to yield the
{\it constraint gradient}, $\nabla D$, a row vector in configuration space.
For a V-F penetration, the constraint gradient expressed in the local indices of the stencil is
\begin{equation}
\nabla D = ({\bf n}^T, -\frac{1}{3} {\bf n}^T, -\frac{1}{3} {\bf n}^T, -\frac{1}{3} {\bf n}^T) .
\label{eqn_post02}
\end{equation}
To elevate the constraint gradient in the $\mathbb{R}^{3n}$ space, we simply map the local indices to their global position and re-write
$\nabla D$ as a $1 \times 3n$ row vector, with zeros everywhere else, in the style of finite element stiffness matrix assembly.
Please note that the inner product $\nabla D {\bf x}$ is also the signed distance, thus $\nabla D {\bf x} = D({\bf x})$.

{\bf Displacement for a single constraint.}
Let unprimed and primed quantities refer to the {\it pre-} and {\it post-}response states, respectively.
For example, ${\bf x}'$ is the post-response position.
A valid response satisfies two properties. First, it pushes colliding objects apart by relocating each of the four vertices along
the constraint direction, {\it i.e.}, ${\bf x}' = {\bf x} + {\bf M}^{-1} \nabla D^T \lambda$ for some (unknown)
non-negative multiplier $\lambda$.
The diagonal mass matrix ${\bf M}$ here is to meet the requirement of conservation law,
thus ${\bf M}^{-1} \nabla D^T \lambda$ is the displacement vector for all four vertices.
Second, the post-response signed distance should be non-negative: $\nabla D' {\bf x}' \geq 0$.

For properly triangulated mesh used in dynamics simulation, assuming equal mass for all vertices is very common.
In this case, together with the carefully chosen $\alpha_{1,2,3} = 1/3$,
the normal ${\bf n}$ of face ${\bf x}_1 {\bf x}_2 {\bf x}_3$ remains unchanged after response.
Thus there is $\nabla D' = \nabla D$ and
it is safe to use $\nabla D {\bf x}'$ as the post-response signed distance.
For an inelastic collision, we seek the maximal dissipation among all valid responses, per the definition of purely inelastic.
Thus we minimize the post-response signed distance:
$\nabla D' {\bf x}' = 0$,
which is equivalent to
$$\nabla D {\bf x}' = \nabla D ({\bf x} + {\bf M}^{-1} \nabla D^T \lambda) = 0, $$
from which $\lambda$ can be computed.

{\bf Displacements for multiple constraints.}
We are typically faced with $k$ penetration constraints, {\it i.e.}, $k$ V-F
penetrations with possibly non-disjoint stencils. Now let $\nabla D = [\nabla D_1^T, \cdots, \nabla D_k^T]^T$ be a $k \times 3n$
matrix whose rows span the possible displacement directions.
For any vector ${\bf \lambda} \in \mathbb{R}^k$, ${\bf x}' = {\bf x} + {\bf M}^{-1} \nabla D^T {\bf \lambda}$
corresponds to the application of a linear combination of correcting displacements to ${\bf x}$. As in the single-constraint case,
we require that $\lambda_1 \cdots \lambda_k$ be non-negative, since displacements are to cancel the negative distance,
and lead to a non-negative post-response distance, {\it i.e.},
that every row of $\nabla D' {\bf x}' = [\nabla D'_1 {\bf x}', \cdots, \nabla D'_k {\bf x}']^T$ be non-negative.

We propose an algorithm for approximating ${\bf \lambda}$ by assuming the post-response distance to be exactly zero:
$\nabla D' {\bf x}' = {\bf 0}$.
As evaluating $\nabla D'$ requires computation of the triangle normals according to their new (unknown) vertex positions,
we instead use $\nabla D$ to approximate $\nabla D'$ and instead demand $\nabla D {\bf x}' = {\bf 0}$.
We relax the conditions on a response being valid to allow
both positive and negative entries in ${\bf \lambda}$,
then ${\bf x}'$ is the minimizer of
\begin{equation}
|| {\bf x}' - {\bf x} ||^2_M, \quad \mbox{subject to } \quad \nabla D {\bf x}' = {\bf 0} .
\label{eqn_minimizer}
\end{equation}
This minimization seeks new vertex positions that are as close as possible to the old positions, in the sense that
position differences are penalized by the corresponding particle masses.
This minimization projects the illegal vertices onto the orthogonal complement of the span of the columns of $\nabla D^T$.
We may repose the above as an extremization of the augmented functional
\begin{equation}
W({\bf x}', {\bf \lambda}) = \frac{1}{2} || {\bf x}' - {\bf x} ||^2_M + (\nabla D {\bf x}')^T{\bf \lambda},
\end{equation}
with respect to $({\bf x}', {\bf \lambda})$, where ${\bf \lambda}$ is a vector of Lagrange multiplier.
The corresponding stationary equations are
\begin{eqnarray}
{\bf 0} = & \frac{\partial W}{\partial {\bf x'}} = & {\bf x'} - {\bf x} + {\bf M}^{-1} \nabla D^T {\bf \lambda}, \\
{\bf 0} = & \frac{\partial W}{\partial {\bf \lambda}} = & \nabla D {\bf x}' .
\end{eqnarray}
The first stationary equation guarantees that the response acts only along the $\nabla D$ direction,
and the second one ensures vanishing negative distance.
Substituting the first into the second yields
\begin{equation}
\nabla D {\bf M}^{-1} \nabla D^T {\bf \lambda} = \nabla D {\bf x}
\label{eqn_sol2}
\end{equation}
for ${\bf \lambda}$, and then substituting it into the first stationary equation to
obtain positional displacement for a set of $k$ simultaneous collisions.

{\bf Elastic collision.}
If we don't have to pursue purely inelastic collision, the post-response signed distance
can be set to certain positive value, i.e., $\nabla D {\bf x}' = {\bf d}$, with ${\bf d} = (d_1, d_2, \cdots, d_k)^T$.
In this case, Equ.~\ref{eqn_sol2} becomes
\begin{equation}
\nabla D {\bf M}^{-1} \nabla D^T {\bf \lambda} = \nabla D {\bf x} - {\bf d} .
\label{eqn_sol3}
\end{equation}
For models that assume isotropic physical properties across the whole mesh,
we can simply set identical values for $d_1$, $d_2$, $\cdots$, $d_k$.

{\bf Implementation.}
After performing DCD on two meshes, we may obtain a number of penetration stencils. To handle these stencils in a
global manner, we follow the concept of {\it impact zone} (IZ) of \cite{bridson02robust,Harmon08robust}.
As each constraint gradient depends on four vertices, an impact zone is defined as a set of
constraint gradients with overlapping stencils. The projection operation (Equ.~\ref{eqn_minimizer})
is performed per IZ, and each IZ corresponds to one linear system of Equ.~\ref{eqn_sol2}.

\section{Diffusion of Displacement Vectors}
\label{sec:vectorfield}

Cloth is often regarded as continuum material, and the dynamics model is designed to
demonstrate corresponding behavior.
In computer simulation particles are inter-connected by springs, and their motion is subject to physics laws.
Thanks to the implicit dynamics solver being used,
the particle velocities form a smoothly varying vector field.
That is to say if two particles are topologically very close, so are their respective velocities.
However when collision constraints enforce position or velocity adjustment, discontinuities will be
introduced to the smooth vector field, causing not only visual artifacts but also simulation instability.
Our solution is to propagate the correcting displacement vectors from the penetrating vertices to the vicinity,
creating a smooth vector field of corrections to be applied globally.
We use the constrained diffusion method given in \cite{Xu2009harmonic,Xu2009texturing}
to compute harmonic vector field.

\section{Applications and Results}
\label{sec:results}

To test the effectiveness of proposed method, we designed several experiments.
In each case at least one mesh is specified as oriented surface,
so the mesh-mesh collision is handled with the proposed method.
The self-collision of each deformable mesh is handled with
the traditional CCD-based method.

The first experiment (Fig.~\ref{fig_teaser}a) simulates a spike (represented as an oriented surface) stabbing a sheet (18,896 vertices),
to show our method's ability to process sharp geometry. At the beginning,
the collision response is disabled and the sheet is pierced by the spike on purpose.
A while later our DCD-based collision response is enabled, and the existing penetration gets resolved
and no new penetration occurs thereafter.

The second experiment (Fig.~\ref{fig_teaser}b) simulates two tori colliding with each other.
Both tori are modeled as oriented surface meshes of 576 vertices.
Air pressure is computed based on its volume and applied to vertices.
We vary the post-response distance ${\bf d}$ in Equ.~\ref{eqn_sol3}
to emulate behaviors from elastic to inelastic collisions.
Note the red torus is not subject to gravity (consider it as aired balloon).

\begin{figure}[!htb]
   \centering
   \includegraphics[width=1.0\linewidth]{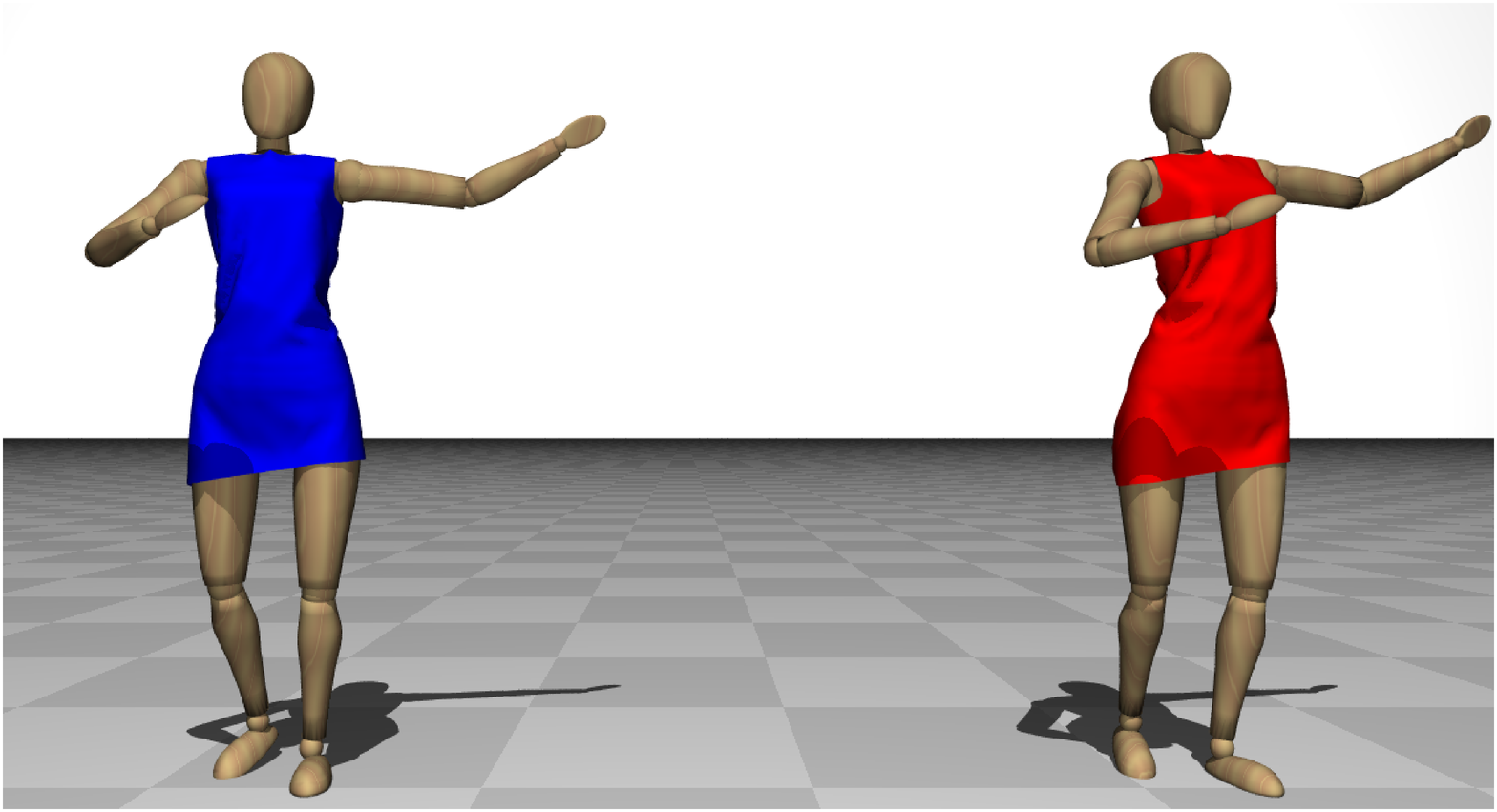}
   \caption{\small By taking a much larger time step to do solid/cloth collision faster,
                   our result (right) does not suffer any quality loss compared to Narain's (left).}
   \label{fig:Img_dressRedBlue}
\end{figure}

In the third experiment (Fig.~\ref{fig:Img_dressRedBlue}),
the proposed method has been integrated into a cloth simulator based on \cite{Narain2012AAR}.
While simulating a dressed dummy, body parts are modeled as oriented surface meshes.
The self-collision of the garment mesh is handled with the simulator's built-in CCD pipeline,
while the body/garment collision is processed with our DCD-based algorithm.
The simulator runs at time step $\Delta t = 0.005s$, and outputs one frame data every eight steps.
To demonstrate that our method tolerates much larger time step than CCD, the body/garment collision
is handled once for every eight steps -- only at the steps when frame data is to be output!
The accompanying video shows there is no quality loss in our simulation result.
Moreover, the new simulator is faster than the old one -- it cuts the overall collision handling time by 40\%.

\section{Discussions and Conclusion}
\label{sec:conclusion}

We have presented a new collision response approach to untangle existing penetrations.
In many circumstances where surface orientation is either explicitly given or implied
in the context (e.g. closed surfaces),
this method is a competitive alternative to the popular CCD-based approach.

Self-healing is the primary advantage of DCD-based response method.
Even if in certain steps there are detected collisions that can not be resolved right away due to various reasons,
they will not ruin the simulation in the following steps.
Of course, the penetration will be processed at earliest possible time.

Another significant benefit of DCD-based approach is the large time step.
As physically based simulation is often computational expensive, acceleration is pursued
in two aspects: speeding-up the per-step computation and using larger time steps.
As new technique \cite{liu13fast} has made it possible for the dynamics solver to take large time step,
the CCD-based collision handling, which is sensitive to the time step, becomes the bottleneck of the simulation.
Our DCD-based approach provides a solution to this issue.
In real production, the artist can opt, in some designated steps, for incomplete response,
as long as the intersecting regions do not expand too much and the rendered images do not suffer from visual artifacts.
In video games, insignificant penetrations are often acceptable as long as they do not create obvious artifacts.

{\bfseries Limitations.}
For the DCD-based response strategy to be widely adopted, it must be extended to be able to handle un-oriented surfaces.
As the algorithm relies on the front/back identification, we must, at least,
be able to dynamically identify the faces of the colliding regions.
This can be achieved either by global intersection analysis \cite{baraff03untangling} or from history information.
In the latter case, the history will only be used to identify front/back, but no cubic collision solver should be involved.

Although diffusion field is created to propagate the positional displacement,
we noticed sometimes the relocated positions introduce self-collisions in the deformable mesh.
It is desirable that resolving one collision does not generate a new one.

\section*{Acknowledgements}

This work was supported by xxxx under the grants \#yyyy.

\bibliographystyle{acmsiggraph}
\bibliography{ch9_bibliography}
\end{document}